\documentclass{amsart}
\usepackage{amssymb,stmaryrd}
\usepackage{amsfonts}
\usepackage{amstext}
\usepackage{algorithmic}
\usepackage{algorithm}
\usepackage{graphicx}
\usepackage{epstopdf}
\usepackage[all]{xy}
\usepackage{MnSymbol}
\usepackage{tikz}

\numberwithin{equation}{section}
\newtheorem{theorem}{Theorem}[section]

\newtheorem{proposition}[theorem]{Proposition}

\theoremstyle{definition}

\newtheorem{example}[theorem]{Example}

\theoremstyle{remark}

\newcommand{\R}{{\mathbb{R}}}

\newcommand{\C}{{\mathbb{C}}}
\newcommand{\Z}{{\mathbb{Z}}}

\newcommand{\F}{{\mathbb{F}}}

\newcommand{\<}{{\langle}}

\renewcommand{\>}{{\rangle}}

\newcommand{\CO}{{\mathcal{O}}}

\newcommand{\CD}{{\mathcal{D}}}

\newcommand{\tens}{\otimes}

\newcommand{\id}{{\rm id}}
\newcommand{\extd}{{\rm d}}
\newcommand{\del}{{\partial}}

\newcommand{\inthom}{{\underline{\rm hom}}}

\newcommand{\nquad}{\kern-10pt}
\begin{document}

\title{Quantum geometry, logic and probability}
\keywords{logic, noncommutative geometry, digital geometry, quantum gravity, duality, power set, Heyting algebra}

\subjclass[2010]{Primary 03G05, 81P10, 58B32,  81R50}

\author{S. Majid}
\address{Queen Mary University of London\\
School of Mathematics, Mile End Rd, London E1 4NS, UK}

\email{s.majid@qmul.ac.uk}
\date{}

\begin{abstract} Quantum geometry on a discrete set means a directed graph with a weight associated to each arrow defining  the quantum metric. However, these `lattice spacing' weights do not have to be independent of the direction of the arrow. We use this greater freedom to give a quantum geometric interpretation of discrete Markov processes with transition probabilities as arrow weights, namely taking the diffusion form $\del_+ f=(-\Delta_\theta+ q-p)f$ for the graph Laplacian $\Delta_\theta$, potential functions $q,p$ built from the probabilities, and  finite difference $\del_+$ in the time direction. Motivated by this new point of view, we introduce a `discrete Schroedinger process' as $\del_+\psi=\imath(-\Delta+V)\psi$ for the Laplacian associated to a bimodule connection such that the discrete evolution is unitary. We solve this explicitly for the 2-state graph, finding a 1-parameter family of such connections and an induced `generalised Markov process' for $f=|\psi|^2$ in which there is an additional source current built from $\psi$.  We also discuss our recent work on the quantum geometry of logic in `digital' form over the field $\F_2=\{0,1\}$, including de Morgan duality and its possible generalisations. 
\end{abstract}
\maketitle 

\section{Introduction}

These are notes growing out of a conference in Krakow with the wonderful question ``Is logic physics?". My view is yes, obviously. Indeed I have long proposed \cite{Ma:pri} to think of Boolean algebras as the `simplest theory of physics', in which case  we should be able to see how physical and geometric issues in more advanced theories emerge from structures already present there.  If so then we can see the origin of physics in the very nature of language and the structure of mathematical and physical discourse, a philosophy that I have espoused as {\em relative realism}.\cite{Ma:pri,Ma:ess,Ma:qg3,Ma:sel} 

One of the things particularly to be explored in this way is the evident role of entropy in the Einstein equations and its deep link with gravity as an ingredient of quantum gravity. My proposal is that while we may not be able to answer this in quantum gravity itself, we can try to move quantum gravity ideas back to simplified settings such as finite graphs\cite{Ma:squ} and Boolean algebras or digital geometry \cite{MaPac2,Ma:boo} to see how they might emerge. The present paper is a step in this direction towards entropy and gravity: we see how probability and irreversible processes can be seen naturally as emerging from quantum geometry. 

Our main new results are in Section~\ref{secmarkov} in the context of graphs, working over $\C$ and $\R$. The idea is to explore a curious feature of quantum geometry applied in the discrete case, namely that the lattice  `square length' $x\to y$ need not be same as for $y\to x$.  In usual lattice geometry we would be focussed on the edge-symmetric case where these coincide. I will argue that the asymmetric generalisation is the natural setting for probability and non-reversible processes. Thus we consider a Markov process 
\[ f^{new}(x)=\sum_{x\to y} p_{x\to y} f(y)+ (1-p(x))f(x),\]
where $p_{x\to y}$ denote the conditional transition probability from $x$ to $y$ and $p(x)=\sum_{y: x\to y} p_{x\to y}$ for each $x$. If we let  $\del_+$ be the finite difference on $\Z$ as the `discrete time' then the Markov process appears as 
\[ \del_+ f= -\Delta f + (q-p)f \]
where $q(y)=\sum_{x:x\to y}p_{x\to y}$ and $\Delta$ is the canonical graph Laplacian associated to  a (typically degenerate) metric inner product defined by the $p_{x\to y}$ where typically indeed the `distance' $x\to y$ need not be the same as for $y\to x$. In fact the quantities that we should really think of  as length  more geometrically are the negative logs $\lambda_{x\to y}$ of the probabilities, and indeed the the shortest length path in  that sense is indeed the path of maximum probability. Moreover, the shortest distance between two points in this sense makes the space into a Lawvere metric space\cite{ApCat}. In our point of view is that this is not so much a `generalisation' as what naturally arises in quantum geometry for a discrete space; the real question is why distances are not direction dependent in the continuum limit. Indeed, it could be argued that many processes are
irreversible so already time between events {\em is} measured in a directed way. In the present context, this is reflected slightly differently in our probabilistic interpretation of `distance' as conditional transition probabilities. 

Finally, our point of view leads us to introduce a `discrete Schroedinger's process'
\[ \del_+\psi=\imath(-\Delta+V)\psi\]
and to ask what connections $\nabla$ are `unitary' in the sense that $f=|\psi|^2$ remains a probability density when we use the associated Laplacian. We find (see Proposition~\ref{propsch}) that the discreteness results in the evolution of $f$ being a generalised Markov process coupled to a probability current source. Indeed, the discrete Schroedinger process being unitary is reversible, so the irreversibility of the Markov process on $f$ is compensated by this source. We construct such a unitary 
Schroedinger process explicitly for a 2-state graph $\bullet\leftarrow\!\to\bullet$, finding a 1-parameter family of suitable connections. Unitary matrices are the essence of quantum computing and what we have done is to construct a certainly family of them quantum geometrically in the spirit of  Schroedinger's equation. Another motivation is the recent formulation of `quantum geodesics'  \cite{Beg:geo,BegMa:geo} which indeed includes the  actual Schroedinger's equation of ordinary quantum mechanics. This work is not immediately applicable  due our use current of discrete time, but these ideas should tie up in future work.

We will use and make reference to a modern constructive `quantum geometry' formalism briefly outlined in Section~\ref{secqg}; see \cite{BegMa} for details and the extensive literature cited therein. This approach starts with a unital algebra $A$ and choice of differential structure $(\Omega^1,\extd)$ on it and then proceeds to metrics, connections, curvature and so forth. Growing out of many years experience with quantum groups but in no way limited to them, it is very different in flavour from the well-known approach of A. Connes \cite{Con} coming out of cyclic cohomology and `spectral triples' as abstract `Dirac operators', though with the possibility of useful interaction between the approaches\cite{BegMa:spe}. 

\begin{figure}\[ \includegraphics[scale=0.36]{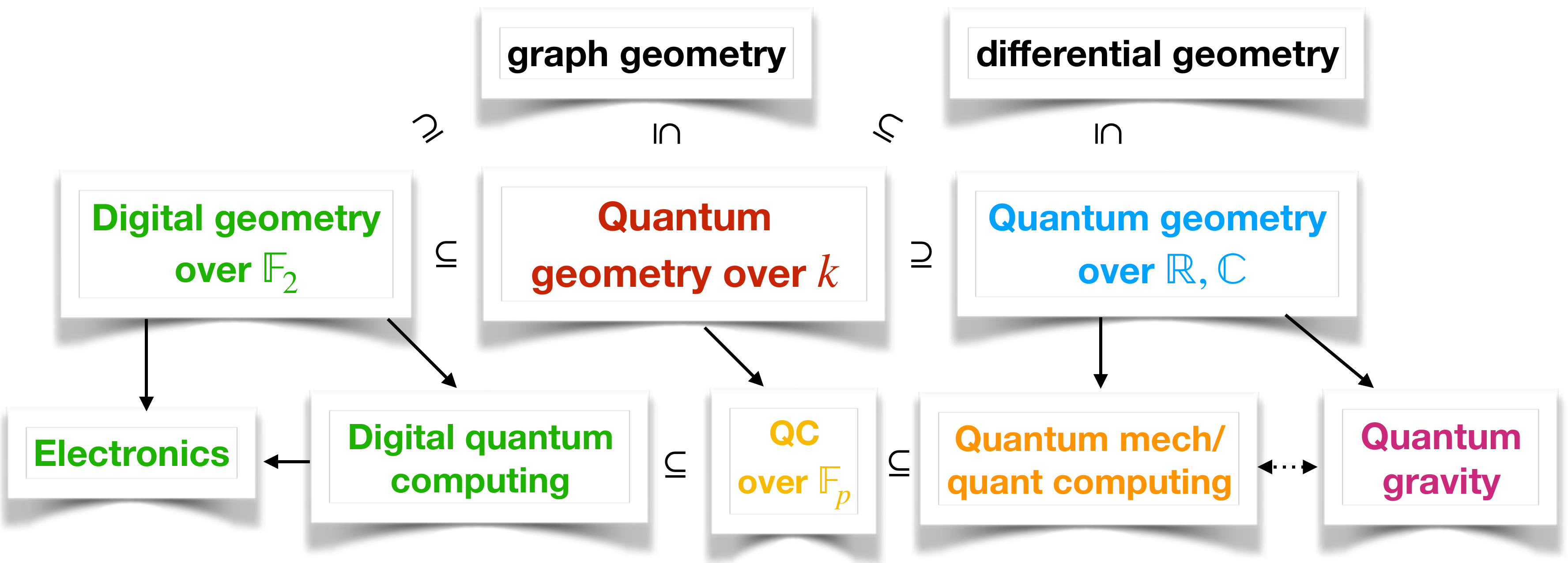}\] \caption{Quantum gravity and quantum mechanics/computing mediated by quantum geometry over $\C$. Digital geometry as a limiting case.\label{figover}}
\end{figure}

Another aspect of the constructive approach is that it works over any field, which makes possible the `digital' case by working over the field $\F_2=\{0,1\}$ of two elements\cite{BasMa, MaPac1,MaPac2} and allows in principle the transfer of geometric ideas to digital electronics, see Figure~\ref{figover}. A brief overview is in Section~\ref{secM2} with some modest new results for the quantum geometry of  $A=M_2(\F_2)$.  In the same vein, one could in theory  put something like a digital wave operator for a black hole background onto a silicon chip. A motivation here is again from quantum computing. While in quantum computing a gate is replaced by a unitary operator (which in turn we  envisage could be constructed  quantum geometrically, e.g. by a  Schroedinger process), the essential feature here is the use of vector spaces (the superposition principle) to massively parallelise computations. However, linear algebra works over any field. Hence if we build our gates quantum geometrically then we could specialise them over other fields, including over $\F_2$ as `digitial quantum computing'. Even if this did not have the speed benefits of actual quantum computing, it would provide conventionally realisable training wheels for the real thing.

\section{Outline of quantum geometry} \label{secqg} 

It is well-known that classical geometry can be formulated equivalently in terms of a suitable algebra of functions on the space. The idea in noncommutative or `quantum' geometry  is to allow this to be any algebra $A$ with identity as our starting point (and now no actual space need exist). We replace the notion of differential structure on a space by specifying a bimodule $\Omega^1$ of differential forms over $A$. A bimodule means we can multiply a `1-form' $\omega\in\Omega^1$ by `functions' $a,b\in A$ either from the left or the right and the two should associate according to 
\begin{equation}\label{bimod} (a\omega)b=a(\omega b).\end{equation}
We also need $\extd:A\to \Omega^1$ an `exterior derivative' obeying reasonable axioms, the most important of which is the Leibniz rule 
\begin{equation}\label{leib} \extd(ab)=(\extd a)b+ a(\extd b)\end{equation}
for all $a,b\in A$. We usually require $\Omega^1$ to extend to forms of higher degree to give a graded algebra  $\Omega=\oplus\Omega^i$ (where associativity extends the bimodule identity (\ref{bimod}) to higher degree). We also require $\extd$ to extend to $\extd:\Omega^i\to \Omega^{i+1}$ obeying a graded-Leibniz rule with respect to the graded product $\wedge$ and $\extd^2=0$. This  `differential structure' is the first choice we have to make in model building once we fixed the algebra $A$. We require that $\Omega$ is then generated by $A,\extd A$ as it would be classically. 

\begin{figure}\[ \includegraphics[scale=0.63]{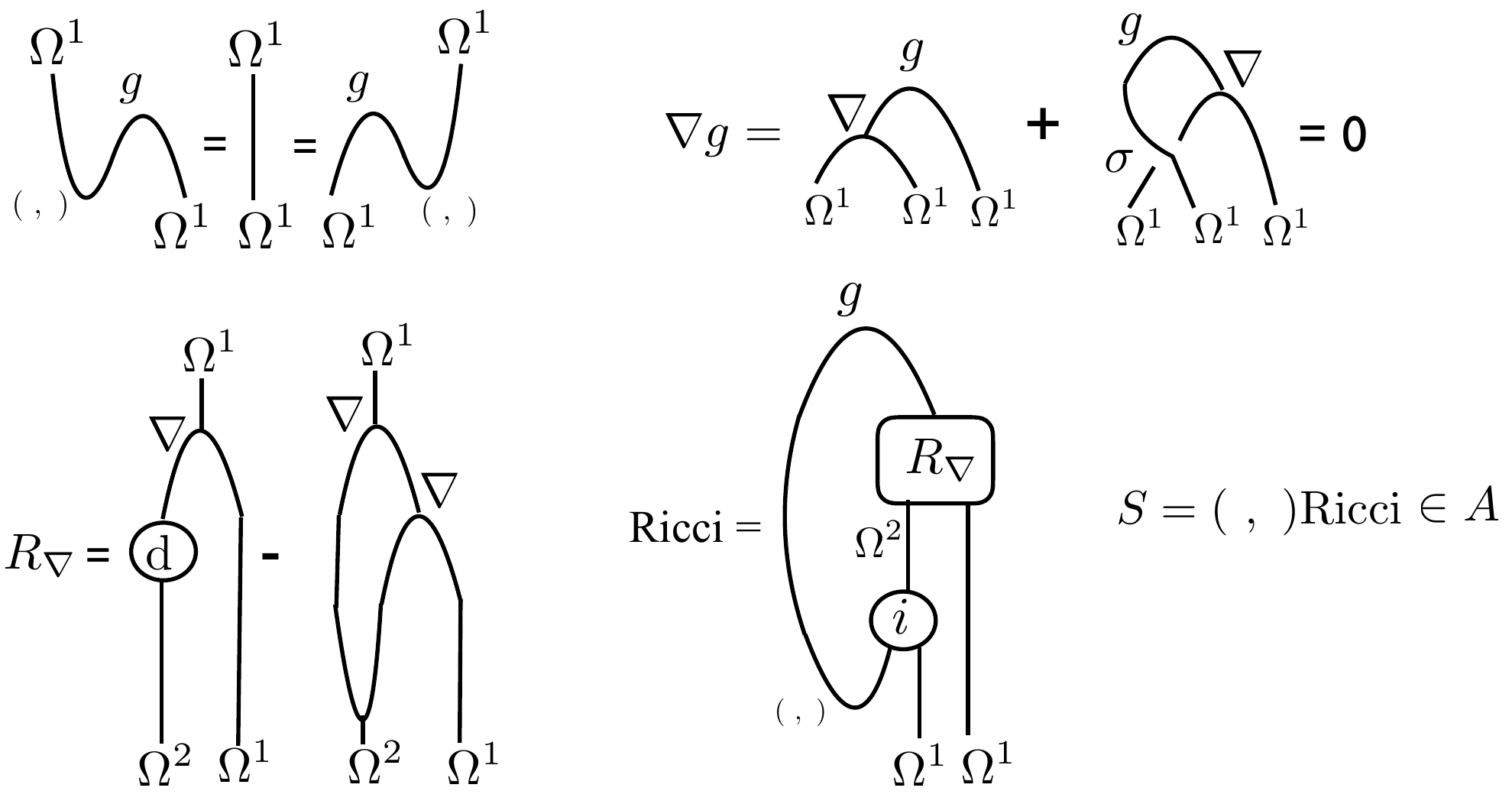}\] \caption{Some axioms of quantum Riemannian geometry. In order: invertibility of a metric, tensor product bimodule connection, Riemann curvature and Ricci curvature tensor and scalar. Diagrams are read down the page as a series of compositions.\label{figaxioms}}
\end{figure}

Next, on an algebra with differential we define a metric as an element $g\in \Omega^1\tens_A\Omega^1$ which is invertible in the sense of a map $(\ ,\ ):\Omega^1\tens_A\Omega^1\to A$ which commutes with the product by $A$ from the left or right and inverts $g$ in the sense
\begin{equation}\label{metricinv}((\omega,\ )\tens_A\id)g=\omega=(\id\tens_A(\ ,\omega))g\end{equation}
 for all 1-forms $\omega$. This is shown in Figure~\ref{figaxioms}.  In the general theory one can require quantum symmetry in the form $\wedge(g)=0$, where we consider the wedge product on 1-forms as a map $\wedge:\Omega^1\tens_A\Omega^1\to A$ and apply this to $g$. However, we don't need to and moreover we can also work with $g$ non-invertible or $(\ ,\ )$ degenerate. 
 
Finally, we need the notion of a connection. A left connection on $\Omega^1$ is a linear map $\nabla :\Omega^1\to \Omega^1\tens_A\Omega^1$ obeying a left-Leibniz rule 
\begin{equation}\label{connleib}\nabla(a\omega)=\extd a\tens_A \omega+ a\nabla \omega\end{equation} 
for all $a\in A, \omega\in \Omega^1$. This might seem mysterious but if we think of a map $X:\Omega^1\to A$ that commutes with the right action by $A$ as a `vector field' then we can evaluate $\nabla$ as a covariant derivative $\nabla_X=(X\tens_A\id)\nabla:\Omega^1\to \Omega^1$ which classically is then a usual covariant derivative on $\Omega^1$. There is a similar notion for a connection on a general `vector bundle' expressed algebraically.  Moreover, when we have both left and right actions of $A$ forming a bimodule, as we do here, we say that a left connection is a {\em bimodule connection}\cite{DV2,BegMa:gra}, if there also exists a bimodule map $\sigma$ such that
\begin{equation}\label{sigma} \sigma:\Omega^1\tens_A\Omega^1\to \Omega^1\tens_A\Omega^1,\quad \nabla(\omega a)=(\nabla\omega)a+\sigma(\omega\tens_A\extd a)\end{equation}
for all $a\in A, \omega\in \Omega^1$.  The map $\sigma$ if it exists is unique, so this is not additional data but a property that some connections have.  The key thing is that bimodule connections extend automatically to tensor products  as 
\begin{equation} \nabla(\omega\tens_A\eta)=\nabla\omega\tens_A\eta+(\sigma(\omega\tens_A(\ ))\tens_A\id)\nabla\eta\end{equation} for all $\omega,\eta\in \Omega^1$, so that metric compatibility now makes sense as $\nabla g=0$. This is shown in Figure~\ref{figaxioms}. A connection is called  QLC or `quantum Levi-Civita' if it is  metric compatible and the torsion also vanishes, which in our language amounts to $\wedge\nabla=\extd$ as equality of maps $\Omega^1\to \Omega^2$. Given a metric inner product $(\ ,\ )$ and a connection $\nabla$, one has divergence and geometric Laplacian
\begin{equation}\label{divlap} \nabla\cdot\omega=(\ ,\ )\nabla\omega,\quad \Delta a=(\  ,\ )\nabla\extd a\end{equation}
for all $\omega\in\Omega^1, a\in A$. 

We also have a Riemannian curvature for any connection,
\begin{equation}\label{curv} R_\nabla=(\extd\tens_A\id-\id\wedge\nabla)\nabla:\Omega^1\to \Omega^2\tens_A\Omega^1,\end{equation} where classically one would interior product the first factor against a pair of vector fields to get an operator on 1-forms. Ricci requires more data and the current state of the art (but probably not the ultimate way) is to introduce a lifting bimodule map $i:\Omega^2\to\Omega^1\tens_A\Omega^1$. Applying this to the left output of $R_\nabla$; we are then free to `contract' by using the metric and inverse metric to define ${\rm Ricci}\in \Omega^1\tens_A\Omega^1$ \cite{BegMa:spe}. This is also shown in Figure~\ref{figaxioms}. 

Finally, and critical for physics, are unitarity or `reality' properties. We mainly work over $\C$ and assume that $A$ is a $*$-algebra (real functions, classically, would  be the self-adjoint elements). We require this to extend to $\Omega$ as a graded-anti-involution (so reversing order with an extra sign when odd degree differential forms are involved) and to commute with $\extd$. `Reality' of the metric and of the connection in the sense of being $*$-preserving are imposed as \cite{BegMa:gra,BegMa:spe}
\begin{equation}\label{realgnab} g^\dagger=g,\quad \nabla\circ *= \sigma\circ\dagger\circ \nabla;\quad (\omega \tens_A\eta)^\dagger=\eta^*\tens_A \omega^*,\end{equation}  where $\dagger$ is a natural $*$-operation on $\Omega^1\tens_A\Omega^1$. These `reality' conditions in a self-adjoint basis (if one exists) and in the classical case would ensure that the metric and connection coefficients  are real.

In the case where there exists $\theta\in \Omega^1$ such that $\extd a=[\theta,a]$, one says that the calculus is {\em inner}. This  is never possible in the classical case but is rather typical in the quantum case. One then has that
any bimodule connection $\nabla$  has the form\cite{Ma:gra}
\[ \nabla \omega= \theta\tens\omega-\sigma(\omega\tens\theta)+\alpha(\omega)\]
for some bimodule maps $\sigma,\alpha$. A canonical (but not classical) choice is $\sigma=\alpha=0$ in which
case
\[ \nabla_\theta= \theta\tens,\quad T_{\nabla_\theta}=-(\ )\wedge\theta,\quad R_{\nabla_\theta}=\theta^2\wedge,\quad \nabla g=\theta\tens g \]
so that this connection cannot be Levi-Civita for a nontrivial exterior algebra or nontrivial metric. Nevertheless, it defines a canonical divergence and canonical Laplacian according to (\ref{divlap}), namely
\[ \nabla_\theta\cdot\omega=(\theta,\omega),\quad \Delta_\theta  a= (\theta,\extd a)= -(\extd a,\theta)+[(\theta,\theta),a]\]
which we will use in Section~\ref{secmarkov} (note that the latter is $\Delta_\theta/2$ in \cite{Ma:gra,BegMa}). Finding actual QLCs is a much more involved problem and usually results in a moduli space of $\nabla$  rather than a unique one.  

\section{Asymmetric discrete geometry and Markov processes}\label{secmarkov}

We are interested in the case $A=k(X)$ of functions on a discrete set $X$ with values in a field $k$. Here we necessarily have $\Omega^1={\rm span}_k\omega_{x\to y}$ with basis labelled by the edges of a graph on $X$. The bimodule structure and exterior derivative are
\[ f.\omega_{x\to y}=f(x)\omega_{x\to y},\quad \omega_{x\to y}.f=f(y)\omega_{x\to y},\quad \extd f=[\theta,f]=\sum_{x\to y} (f(y)-f(x))\omega_{x\to y},\]
where $\theta=\sum_{x\to y}\omega_{x\to y}$. By definition, we don't include self-arrows in the graph (but it can be useful to {\em extend} the graph to allow them). We assume that our graph is {\em bidirected} in the sense that if $x\to y$ is an arrow then so is $y\to x$. In this case we can define a {\em metric inner product} $(\ ,\ ):\Omega^1\tens_A\Omega^1\to A$ as the bimodule map
\[ (\omega_{x'\to y'},\omega_{y\to x})=\delta_{x',x}\delta_{y',y}p_{y\to x}\delta_x,\]
for some metric weights $p_{y\to x}$. {\em Unlike usual quantum geometry} we now do {\em not} suppose nondegeneracy in the sense that these are all nonzero. The arrows $x\to y$ for which $p_{x\to y}\ne 0$ are the more relevant subgraph but it is convenient to use the full bidirected graph for $\Omega^1$ with the price that that some of the weights could vanish. Given the bimodule inner product, we use the canonical graph Laplacian\cite{Ma:gra} as above, which works out as
\begin{equation}\label{delth} -\Delta_\theta f=(\extd f,\theta)=\sum_{x\to y} (f(y)-f(x))\delta_x p_{y\to x}.\end{equation}

For the moment we work over $\R$ and define functions
\[  p,q\in \R(X),\quad  p(x)=\sum_{y:x\to y} p_{x\to y},\quad q(x)=(\theta,\theta)=\sum_{y:x\to y} p_{y\to x},\quad  \forall x,y\in X.\]
We  will assume that $(\ , \ )$ is {\em stochastic} in the sense that
\[ p_{x\to y}\ge 0,\quad \forall x\to y,\quad p(x)\le 1,\quad \forall x\in X.\]
In other words, we do noncommutative geometry but with weights in the Heyting algebra $[0,1]$. In more conventional terms, we define a Markov transition matrix by $P_{x,y}=p_{x\to y}$ if $x\to y$ and $P_{x,x}=1-p(x)$ with other entries zero, which is then right stochastic (all rows sum to 1). One could equally extend the graph to allow self-arrows with $p_{x\to x}=1-p(x)$ on the extended graph, defining a generalised calculus $\tilde\Omega^1$ in the sense of \cite{MaTao:dua}. 

We also consider $f\in \R(X)$ a probability distribution so that $f(x)$ is the probability for each event $x$, or in vector terms $f=(f(x))$ is a stochastic row vector i.e. its elements are $\ge 0$ and sum to 1.  A Markov process with transition probabilities $p_{x\to y}$  is an evolution of such distributions, i.e. labelled by a step index $i$, with
\begin{equation}\label{marf} f_{i+1}(x)= \sum_y f_i(y) P_{y,x}= \sum_{y} f_i(y) p_{y\to x}+ (1-p(x)) f_i(x)\end{equation}
with the convention that $p_{y\to x}=0$ if $y\to x$ is not an arrow of the active graph.  Here $f_{i+1}$ is again stochastic since $\sum_x f_{i+1}(x)=\sum_yf_i(y)$ so that the normalisation is preserved. 

The other ingredient is that the lattice line $\Z$ can be viewed as a graph $\cdots\bullet_i\to \bullet_{i+1}\to\cdots$ and as such there is a 1-dimensional differential calculus $\Omega^1(\Z)$ defined by the graph. As it happens, this is a Cayley graph associated to the additive generator $1$, which means there is a basic left-invariant 1-form $e_+$ with bimodule relations $e_+ f=R_+(f)e$ where $R_+(f)_i=f_{i+1}$ and exterior derivative $\extd f=(\del_+ f)e_+$,  where $(\del_+f)_i=f_{i+1}-f_i$ is the usual 1-step discrete time derivative. This is not bidirected and not suitable for a $*$-calculus (we would be need $\Omega^1$ to be 2-dimensional as in  \cite{Ma:haw}) but is more relevant at the moment. 

\begin{proposition}\label{propmar} A discrete Markov process on a time-dependent probability distribution $f_i(x)$ on $X$ appears naturally as the quantum differential equation
\[ \del_+f= -\Delta_\theta f+ (q-p) f\] 
where $\del_+$ acts on the time variable $i$, $\extd$ on the right is for the graph calculus  on $X$ and $(\ ,\ )$, $\Delta_\theta$ are defined by the transition probabilities.
\end{proposition}
\proof We have already done the work and it remains only to write (\ref{marf}) in terms of $\Delta_\theta$ in (\ref{delth}).  \endproof

\begin{figure}
\[ \includegraphics[scale=0.6]{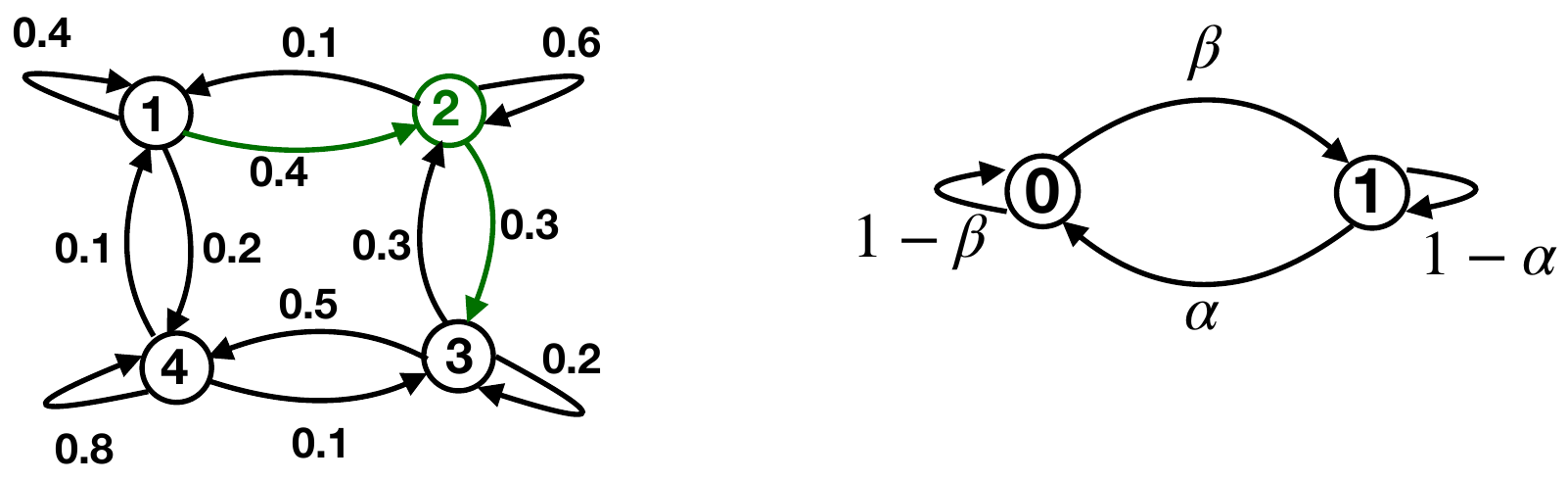}\]
\caption{Markov process state diagrams on vertex set $X$ with transition probabilities labelling arrows of an extended graph. Associated functions on the left are $p=(0.6,0.4,0.8,0.2), q=(0.2,0.7,0.4, 0.7)$ in vertex order as numbered.  
\label{figmarkov}}
\end{figure}

Thus a Markov process is nothing other than the noncommutative diffusion equation with a certain potential term which vanishes in the doubly stochastic case where $p=q$. An example of a right stochastic matrix is shown in Figure~\ref{figmarkov}. The novel idea here is that probabilities play the role of `quantum metric inner product' and that we obtain a quantum geometric picture if we use the canonical Laplacian associated to this. The latter can coincide with the Laplacian for other more geometric connections (there are several examples in \cite{BegMa} where that happens). 

Inspired by this, we can still see we can extend the geometric meaning of a Markov process by tropicalisation, i.e., by writing 
\[ p_{x\to y}= e^{-\lambda_{x\to y}},\quad  p(x)=e^{-\lambda_{x\to x}},\quad  \lambda_{x\to y}, \lambda_{x\to x}\ge 0\]
and observe that the transition probability after $n$-steps is
\[ P^n_{x,y}=\sum_{\gamma: x\to \cdots\to y} e^{- \lambda(\gamma)};\quad \lambda(\gamma)=\sum_{i=0}^{n-1}\lambda_{x_i\to x_{i+1}}\]
where $\gamma=x_0\to x_1\cdots \to x_{n-1}\to x_n$ is an $n$-step path  from $x_0=x$ to $x_n=y$ (in the extended graph where we allow self-steps). If we think of $\lambda_{x\to y}$ as some kind of more geometric actual length associated to the arrow or self-arrow then $\lambda(\gamma)$ is the length of the path $\gamma$. Thus we see how a generalised form of Riemannian geometry emerges as an interpretation of a Markov process, or conversely how Markov processes emerge naturally from generalised Riemannian geometry, with a contribution of maximal probability equated to a path of shortest length. It is generalised in the sense that `lengths' are direction dependent; there is no assumption that $\lambda_{x\to y}=\lambda_{y\to x}$ or even that one is nonzero when the other is,  and we also allow self-arrow lengths $\lambda_{x\to x}$. Figure~\ref{figmarkov} shows the shortest path in this sense between 1 and 3  as via 2. In fact the generalised geometry here
is equivalent to the notion of a Lawvere metric space\cite{ApCat} on the vertex set, where $d(x,y)$ is the length of the shortest path. The main difference is that in that context one would have $\lambda_{x,x}=0$ whereas in our case  these are determined by the values on the arrows of the unextended graph. But in both cases the self-arrow carries no new information and the data is the same. Moreover, this difference does not affect the shortest length provided these are all $\ge 0$, which {\em is} a restriction on the unextended graph weights in our case.  In other words probability is {\em not quite} freely labelling the arrows by lengths $\lambda_{x\to y}$, there is a restriction
\[ \sum_{y: x\to y} e^{-\lambda_{x\to y}}\le 1\]
at every vertex. This has the character of some kind of lower bound on the `distances' from every vertex except that it is not one minimum e.g. Planck length, but shared between all local directions. Our  observations may also be related to path integration defined by stochastic processes, even though our starting point is a fresh one.

Finally, we consider if there is a Schroedinger evolution version of a Markov process. We recall that in usual quantum mechanics, if $\psi$ obeys $\dot\psi= ({\imath\hbar\over 2 m}\Delta+ {V\over \imath\hbar})\psi$ and normalisation $||\psi||_{L^2}=1$ then the probability density $f=|\psi|^2$ obeys the conservation law
\[ \dot f+ \nabla\cdot J=0,\quad J=-{\imath\hbar\over 2m}(\bar\psi\nabla\psi - \psi \nabla\bar\psi)\]
where $J$ is the probability current. This implies that $\int \dot f=0$ so that $\int f=1$ holds at all times. Also note the formal similarity of the Schroedinger equation to an  `imaginary time' (Wick rotation) version of the diffusion one.

We first compute in general that if we are on an inner $*$-differential algebra  with  $V=V^*\in A$, $\theta^*=-\theta$ and $*(\ ,\ )={\rm flip}(*\tens *)$ and set $f=\psi^*\psi$ as dependent on an additional continuous time with 
\[ \dot\psi=\imath(-\Delta_\theta+ V)\psi=\imath (\extd\psi,\theta)+\imath V\psi=-\imath(\theta,\extd\psi)+\imath[(\theta,\theta),\psi]+\imath V\psi\]
then
\[ \dot \psi^*=\imath(\theta,\extd \psi^*)-\imath \psi^*V\]
so that 
\[ \dot f=\dot\psi^*\psi+\psi^*\dot\psi=\imath(\theta,\extd \psi^*)\psi-\imath\psi^*(\theta,\extd\psi)+\imath\psi^*[(\theta,\theta),\psi].\]
If $A$ is commutative then this reduces to 
\[ \dot f=-(\ ,\ )(\theta\tens J)=-\nabla_\theta\cdot J,\quad  J=\imath\left((\extd\psi)\psi^*-(\extd\psi^*)\psi\right)\]
as the probability current 1-form, where $\nabla_\theta=\theta\tens$ is the canonical  connection. This remark only applies to continuous time, and moreover we do not necessarily have $\int:A\to \R$ specified, and even if we did, we may not have $\int \nabla_\theta\cdot J=0$ for all $\psi$. In case of an honest quantum metric and QLC, we might expect that $\int$ can be chosen depending on the metric so that integral of a divergence does vanish as in Riemannian geometry (it is not clear how generally we could do this). However, in our asymmetric setting, this seems unlikely both mathematically and physically; we should not expect actual conserved probability but rather we have a formal similarity as well as an element of irreversibility. A partial result at our graph theory level is the following.

\begin{proposition}\label{propsch} Let $\psi_i\in \C(X)$ be complex valued, $V\in \R(X)$ real valued, $f_i=|\psi_i|^2$ and suppose that
\[ \del_+\psi = \imath(-\Delta_\theta + V)\psi, \]
where $\del_+$ acts on the discrete time index $i$. Then
\[ \del_+ f= V(-\Delta_\theta +V)f+ V(\extd\bar\psi,\extd \psi)-\nabla_\theta\cdot J\]
\[ J=\imath((\extd\psi)\bar\psi-(\extd\bar\psi)\psi)-{1\over 2}((\extd\psi)\Delta_\theta\bar\psi+(\extd\bar\psi)\Delta_\theta\psi)\]
which we can also write as
\[\del_+ f = V^2 f-\nabla_\theta\cdot J_V\]
\[ J_V=\imath((\extd\psi)\bar\psi-(\extd\bar\psi)\psi)+(\extd\psi)(-{1\over 2}\Delta_\theta+V)\bar\psi+(\extd\bar\psi)(-{1\over 2}\Delta_\theta+V)\psi\]
\end{proposition}
\proof Here  $-\Delta_\theta \psi=(\extd\psi,\theta)$  etc., so 
\[ \psi^{new}=\psi+ \imath((\extd\psi,\theta)+V\psi),\quad \bar\psi^{new}=\bar\psi- \imath((\extd\bar\psi,\theta)+V\bar\psi)\]
\begin{align*}
f^{\rm new}&=(\bar\psi- \imath((\extd\bar\psi,\theta)+V\bar\psi))(\psi+ \imath((\extd\psi,\theta)+V\psi))\\
&=f+ ((\extd\bar\psi,\theta)+V\bar\psi)((\extd\psi,\theta)+V\psi) + \imath \bar\psi (\extd\psi,\theta)- \imath(\extd\bar\psi,\theta)\psi\\
&=(1+V^2)f+ (\extd\bar\psi,\theta)(\extd\psi,\theta)+ (\extd\psi,\theta)(\imath+ V)\bar\psi+ (\extd\bar\psi,\theta)(-\imath+V)\psi
\end{align*}
with a cross term $V\bar\psi\psi$ cancelling. Next
\[ (\extd \bar\psi,\theta)\psi=(\extd\bar\psi,\theta\psi)=(\extd\bar\psi,\extd\psi)+(\extd\bar\psi,\psi \theta)= (\extd\bar\psi,\extd\psi)+((\extd\bar\psi)\psi, \theta)\]
so this combined with $\bar\psi(\extd \psi,\theta)=(\bar\psi\extd \psi,\theta)$ contributes $V(\extd f,\theta)+ V(\extd\bar\psi,\extd\psi)$ to $f^{new}$. We also note that
\[- (\theta, (\extd\psi)\bar\psi- (\extd\bar\psi)\psi)= - (\theta,\extd\psi)\bar\psi+(\theta,\extd\bar\psi)\psi=(\extd\psi,\theta)\bar\psi-(\extd\bar\psi,\theta)\psi\]
which times $\imath$ verifies the first term of $J$. It remains to note that
\[ -(\theta,(\extd\bar\psi)(\extd\psi,\theta))=-(\theta,(\extd\psi)(\extd\bar\psi,\theta))=(\extd\bar\psi,\theta)(\extd\psi,\theta)\]
which times 1/2 for each expression as in $J$ gives the remaining term needed for $f^{new}$. This gives our first expression for $\del_+f=f^{new}-f$.  For the second expression we note that
\[ -(\theta, (\extd\bar\psi)V\psi+(\extd\psi)V\bar\psi)=V\bar\psi(\extd\psi,\theta)+V\psi(\extd\bar\psi,\theta)=V(\extd f,\theta)+V(\extd\bar\psi,\extd\psi)\]
as required for the extra terms in $J_V$ to replace the corresponding terms in $f^{new}$. \endproof

Writing 
\[ J=\sum_{x\to y}J_{x\to y}\omega_{x\to y},\quad \tilde\psi(y)=\sum_{y:x\to y}\psi(y)p_{y\to x},\]
explicit formulae for the various terms are
\[ (\extd\bar\psi,\extd\psi)(x)=-\sum_{y:x\to y}f(y)p_{y\to x}- f(x)q(x)+\sum_{y:x\to y}(\psi(x)\bar\psi(y)+\bar\psi(x)\psi(y))p_{y\to x}\]
\begin{align*} J_{x\to y}=&\imath(\psi(y)\bar\psi(x)- \bar\psi(y)\psi(x))+ {1\over 2}\Big(\psi(y)\bar{\tilde\psi}(x)+\bar\psi(y)\tilde\psi(x) -\psi(x)\bar{\tilde\psi}(x)-\bar\psi(x)\tilde\psi(x)\\
&-(\psi(y)\bar\psi(x)+\bar\psi(y)\psi(x))q(x)\Big)+ f(x)q(x)\end{align*}
and the additional terms in $J_V{}_{x\to y}$ are
\[  V(x)(\psi(y)\bar\psi(x)+\bar\psi(y)\psi(x)-2f(x))  \]
A direct calculation from $f^{new}$ in the proof above also gives:

\begin{proposition} For the discrete Schroedinger evolution $\del_+\psi=\imath(-\Delta_\theta+V)\psi$, we have 
\[ \sum_X \del_+ f=\sum_X\left((V-q)^2 f+  (V-q)(\bar\psi\tilde\psi+\bar{\tilde{\psi}}\psi)+  |\tilde\psi|^2+\imath(\bar\psi\tilde\psi-\bar{\tilde{\psi}}\psi)\right)\]
for  $f=|\psi|^2$. 
\end{proposition}

This is typically not zero, i.e.  the discrete Schroedinger evolution with the $\nabla_\theta$ connection does not leave the $l^2$-norm $\sum_x |\psi|^2$ constant,  i.e. does not consist of unitary steps. This is visible even for $V=q$, when $\del_+\psi=\imath\tilde\psi$, and is due in part to the 1-sided step $\del_+f$  not reflecting a $*$-calculus on $\Z$. It is also due to the connection $\nabla_\theta$ being a convenient but not necessarily physical choice. Nevertheless, we see from the proposition that $\del_+f$ has a certain form which, when $V=0$, is $\del_+ f=-\nabla_\theta\cdot J$  as expected in quantum mechanics from a formal point of view,  while for other $V$ also contains a Markov-process like element.  

More generally, we should consider the above with other bimodule connections $\nabla$ on $\Omega^1$ and the associated $\Delta$ and $\nabla\cdot$ from (\ref{divlap}). We will say that a connection $\nabla$ is {\em unitary} if there exists a potential $V$ such that $\del_+\psi =\imath(-\Delta+V)\psi$ indeed has unitary steps so that the total probability is conserved. In the remainder of this section, we explore this idea for the simplest example of a two state process. 

\begin{example} Let $X=\{0,1\}$ with $\alpha=p_{1\to 0}>0$ and $\beta=p_{0\to 1}>0$ define a classical 2-state Markov process  as shown on the right in Figure~\ref{figmarkov}. In addition to this process interpreted as in Proposition~\ref{propmar} in terms of $\nabla_\theta$,  we consider the same ideas as in Proposition~\ref{propsch}  but for Laplacian and divergence given by a general bimodule connection $\nabla$.  It is known from \cite[Lemma~2.1]{Ma:squ} that general these take the form
\[ \nabla\theta= (1-b)\theta\tens\theta,\quad b(0)=s,\quad b(1)=t\]
for two complex parameters $s,t$. Here, $\theta= \omega_{0\to 1}+\omega_{1\to 0}$ is a basis over the algebra so it is enough to specify $\nabla$ on this. Its general value deduced from the Leibniz rule
\[ \nabla(\psi\theta)=\extd \psi\tens\theta+\psi\nabla\theta;\quad \extd \psi=(\psi(1)-\psi(0))(\omega_{0\to 1}-\omega_{1\to 0})\]
comes out as
\begin{equation}\label{nablas} \nabla\omega_{0\to 1}=\omega_{1\to 0}\tens\omega_{0\to 1}-s\omega_{0\to 1}\tens\omega_{1\to 0},\quad \nabla\omega_{1\to 0}=\omega_{0\to 1}\tens\omega_{1\to 0}-t\omega_{1\to 0}\tens\omega_{0\to 1}.\end{equation}
Meanwhile, the metric inner product is
\[ (\omega_{0\to 1},\omega_{1\to 0})=\alpha\delta_0,\quad (\omega_{1\to 0},\omega_{0\to 1})=\beta\delta_1\]
and it is known also from \cite[Lemma~2.1]{Ma:squ}  that $\nabla$ is metric compatible (and hence a QLC for the standard exterior algebra)  if and only if $\beta=\pm\alpha$, which in our context means $\beta=\alpha$, and  $t=s^{-1}$. Moreover, in this case it is $*$-preserving if and only if $s$ has modulus 1. So there is a circle of $*$-preserving QLCs, including the obvious $s=t=-1$ with $\nabla\theta=2\theta\tens\theta$. By contrast, the canonical one we studied above was $\nabla_\theta \theta=\theta\tens\theta$ at $s=t=0$. 

We proceed with a general bimodule connection (\ref{nablas}) for our discussion, remembering that it is a QLC only when $\beta=\alpha$ and $s=t^{-1}$. The metric inner product is
\[ (\omega_{0\to 1},\omega_{1\to 0})=\alpha\delta_0,\quad (\omega_{1\to 0},\omega_{0\to 1})=\beta\delta_1\]
so that the Laplacian comes out as as
\[ (\Delta\psi)(0)=\delta_\psi \beta(1+t),\quad (\Delta\psi)(1)=-\delta_\psi\alpha(1+s);\quad \delta_\psi:=\psi(1)-\psi(0)\]
The discrete Schroedinger evolution $\del_+\psi=\imath(-\Delta+V)\psi$ then corresponds to the matrix step on the column vector $\psi(0),\psi(1)$,
\[ \psi^{new}=U\psi;\quad U= \begin{pmatrix} 1+\imath V(0) -\imath\alpha(1+s) & \imath\alpha(1+s)\\ \imath\beta(1+t) & 1+ \imath V(1) - \imath\beta(1+t)\end{pmatrix},\]
which is unitary for generic parameters if and only if  
\[ \imath\alpha(1+s)=-e^{\imath\phi}\bar z,\quad |1+\imath V(1) - z |^2+ |z|^2=1,
\quad 1+ \imath V(0)=-e^{\imath\phi}(2 \bar z  + 1-\imath V(1))  \]
where $z=\imath\beta(1+t)$ and $e^{\imath\phi}$ is some phase. This has solution 
\begin{equation}\label{solphi}V(0)=V(1)=0,\quad z={1\over 2}(1-e^{\imath\phi}),\quad s=-1-{\imath\over 2\alpha}(1-e^{\imath\phi}),\quad  t=-1-{\imath\over 2\beta}(1-e^{\imath\phi})\end{equation}
for a free angle parameter $\phi$, with resulting Schroedinger evolution step
\begin{equation}\label{Uphi} U=\begin{pmatrix}
 1-z & z \\
 z& 1-z \\
\end{pmatrix}
=e^{\imath\phi\over 2}\begin{pmatrix}
 \cos({\phi\over 2}) & -\imath\sin({\phi\over 2})\\ -\imath\sin({\phi\over 2}) & \cos({\phi\over 2}) 
\end{pmatrix}.\end{equation}
{\em Thus the 2-point graph has a 1-parameter circle of bimodule connections $\nabla$ which are `unitary' in the sense that the step operator $U$ is unitary.} Some examples are:

(i) $\phi=0$ or $z=0$ gives $s=t=-1$ in the QLC family and
\[ \nabla\theta=2\theta\tens\theta,\quad U=\id,\]

(ii) $\phi=0$ or $z=1$   gives $s=-1-{\imath\over 2\alpha}$ and $t=-1-{\imath\over 2\beta}$ and
\[ \nabla\theta=(2+{\imath \over 2\gamma})\theta\tens\theta;\quad \gamma(0)=\alpha,\quad \gamma(1)=\beta;\quad U=\begin{pmatrix}0&1\\1&0\end{pmatrix}.\]
By contrast, our canonical $\nabla\theta=\theta\tens\theta$ at $s=t=0$  is not on this circle. 

\begin{proposition} For the 1-parameter circle of unitary discrete Schroedinger evolutions (\ref{solphi})--(\ref{Uphi}),  and writing $f=|\psi|^2$ in vector notation, we find
\[ f^{new}=\begin{pmatrix} \cos^2({\phi\over 2}) & \sin^2({\phi\over 2})\\ \sin^2({\phi\over 2}) & \cos^2({\phi\over 2})\end{pmatrix}f +{\imath\over 2}\sin(\phi) \det(\bar\psi\tens \psi)\begin{pmatrix}1\\ -1\end{pmatrix},\]
where $\det(\bar\psi\tens\psi)=\bar\psi(0)\psi(1)-\bar\psi(1)\psi(0)$. 
The first term is a general left and right stochastic Markov process on $f$. In quantum geometric terms,
\[ \del_+\psi= -\imath\Delta\psi,\quad \del_+f={1\over 2\imath}(1-e^{-\imath\phi})\Delta f- \nabla\cdot J,\]
\[  J={1\over 2\imath} (1+e^{-\imath\phi}) \det(\bar\psi\tens \psi)(\omega_{01}-\omega_{10}). \]
\end{proposition}
\proof Here 
\begin{align*} f^{new}(0)&=|\psi^{new}(0)|^2=|(1-z)\psi(0)+z\psi(1)|^2\\
&=|1-z|^2f(0)+|z|^2f(1)+(1-\bar z)z\bar\psi(0)\psi(1)+ \bar z(1-z)\bar\psi(1)\psi(0)\\
&=\sin^2({\phi\over 2})f(0)+ \cos^2({\phi\over 2})f(1)+ {\imath\over 2}\sin(\theta)(\bar\psi(0)\psi(1)-\bar\psi(1)\psi(0))\end{align*}
and similarly for $f^{new}(1)=|\psi^{new}(1)|^2=|(1-z)\psi(1)+z\psi(0)|^2$. Note that the entries of $f$ remain positive and summing to 1 even though there is an extra divergence term. 

We next consider a general 1-form $J=J_{01}\omega_{0\to 1}+J_{10}\omega_{1\to 0}$ for  constants $J_{01},J_{10}$ and find from (\ref{nablas}) and (\ref{solphi}) that
\begin{align*}\nabla\cdot J&= (\ ,\ )\big(J_{01}(\omega_{1\to 0}\tens\omega_{0\to 1}- s\omega_{0\to 1}\tens\omega_{1\to 0})+ J_{10}(\omega_{0\to 1}\tens\omega_{1\to 0}- t\omega_{1\to 0}\tens\omega_{0\to 1})\big)\\
&=(J_{10}-sJ_{01})\alpha \delta_0+ (J_{01}-t J_{10})\beta\delta_1=(J_{01}+J_{10})\gamma+ {\imath\over 2}(1-e^{\imath\phi})j\end{align*}
where $\gamma(0)=\alpha, \gamma(1)=\beta$ and $j(0)=J_{01}, j(1)=J_{10}$ are functions on $X$. If we set $J_{01}=-J_{10}={1\over 2\imath}(1+e^{-\imath\phi})\det(\bar\psi\tens\psi)$ as stated then $-\nabla\cdot J$ gives the second term of $f^{new}$ as required. 

Moreover, $\extd f=(f(1)-f(0))(\omega_{0\to 1}-\omega_{1\to 0})$. Hence using the above computation, 
\[ \Delta f=\nabla\cdot\extd f={\imath\over 2}(1-e^{\imath\phi})(f(1)-f(0))\begin{pmatrix}1\\-1\end{pmatrix}={\imath\over 2}(1-e^{\imath\phi})\begin{pmatrix}-1 & 1\\ 1 & -1\end{pmatrix}f\]
in vector notation for $f$, which recovers the Schroedinger evolution $U=1-\imath\Delta$ in (\ref{Uphi}) as expected. Moreover, the stated first term of $\del_+f$ is then
\[ {1\over 2\imath}(1-e^{-\imath\phi}){\imath\over 2}(1-e^{\imath\phi})\begin{pmatrix}-1 & 1\\ 1 & -1\end{pmatrix}f=\sin^2({\phi\over 2})\begin{pmatrix}-1 & 1\\ 1 & -1\end{pmatrix}f\]
as required in $f^{new}-f$.  \endproof \end{example}

For the $l^2$-norm, we took the constant measure in summing over $X$. One could also, in principle, introduce a measure related to the quantum metric $(\ ,\ )$. This would be more in keeping with Riemannian geometry but is not so natural from our point of view on Markov process (where the usual constant measure is preserved). 

\section{Digital geometry of  $2\times 2$ matrices} \label{secM2}

Quantum Riemannian geometry and quantum gravity on one of the simplest graphs, a quadrilateral,  was recently achieved \cite{Ma:squ} (with Lorentzian style negative square-length  weights on two of the sides). Here we briefly look at the complementary example of a noncommutative finite geometry, namely the humble algebra of $2\times 2$ matrices $M_2(\C)$. Its quantum Riemannian geometry turns out to be rather rich and is  not fully explored. 

We take the standard 2D $*$-differential calculus from \cite{BegMa:spe, BegMa} with a basis of central 1-forms $s,t\in \Omega^1(M_2)$, differential and exterior algebra
\[\extd a= [E_{12},a]s+ [E_{21},a]t,\quad   s^2=t^2=0,\quad s^*=-t,\quad s\wedge t=t\wedge s,\quad \extd s=2\theta\wedge s,\quad\extd t=2\theta\wedge t,\]
which is inner with $\theta=E_{12}s+E_{21}t$. Here $E_{ij}$ is the matrix with $1$ in the $(i,j)$ place and 0 elsewhere. 

Next, as the basis is central and the metric has to be central to be invertible in the bimodule sent, any invertible $2\times 2$ complex matrix $g_{ij}$ in our basis can be taken as metric coefficients, with the condition that $g_{12}=-g_{21}$ if we want to impose quantum symmetry, and $g_{22}=\overline{g_{11}}$, $g_{12}$ real if we want $g$ to be `real' in the required hermitian sense. It is easy to see that
\begin{equation}\label{M2canQLC} \nabla s=2\theta\tens s,\quad  \nabla t=2\theta\tens t,\quad \sigma=-{\rm flip},\quad R_\nabla=0\end{equation}
on the generators is a flat QLC simultaneously for all quantum metrics. But there are typically many more QLCs. The actual moduli of these has only been computed in  \cite[Example~8.13 and Exercise 8.3]{BegMa} for a couple of sample quantum metrics $g_1,g_2$, with results there as follows. 

(i) $g_1=s\tens t-t\tens s$. This has a principal 4-parameter moduli of QLCs of the form
\[ \nabla s=2\theta\tens s - \begin{pmatrix}0&\mu\alpha\\ \beta & 0\end{pmatrix}s\tens s -\begin{pmatrix}0&\alpha\\ \nu\beta & 0\end{pmatrix}g_1 +\begin{pmatrix}0&\nu\alpha\\ \nu^2\beta+(\mu\nu-1)\alpha & 0\end{pmatrix}t\tens t \]
\[ \nabla t=2\theta\tens t+\begin{pmatrix}0&\mu^2\alpha+(\mu\nu-1)\beta\\ \mu\beta & 0\end{pmatrix}s\tens s+\begin{pmatrix}0&\mu\alpha\\ \beta & 0\end{pmatrix}g_1- \begin{pmatrix}0&\alpha\\ \nu\beta & 0\end{pmatrix} t\tens t.\]
The $\alpha=\beta=\mu=\nu=0$ point is the flat one (\ref{M2canQLC}). We rescaled the $\mu,\nu$ compared to \cite{BegMa} here and in the next case. 

(ii) $g_2=s\tens s+ t\tens t$. This has a principal 3-parameter moduli of QLCs of the form
\[  \nabla s=2E_{21}t\tens s + \begin{pmatrix}0&\mu\rho\\ 2\mu-\rho(1+\mu(\mu+\nu)) & 0\end{pmatrix}s\tens s +\begin{pmatrix}0&-\rho \\ \mu\rho & 0\end{pmatrix}g_1 +\begin{pmatrix}0&\nu\rho\\ \rho & 0\end{pmatrix}t\tens t \]
\[\nabla t=2E_{12}s\tens t +\begin{pmatrix}0&- \rho\\ \mu\rho & 0\end{pmatrix} s\tens s -\begin{pmatrix}0&\nu \rho \\ \rho & 0\end{pmatrix}g_1 +\begin{pmatrix}0& -2\nu+\rho(1+\nu(\mu+\nu))\\ \nu\rho & 0\end{pmatrix}t\tens t. \]
There are restrictions on the parameter over $\C$ for a $*$-preserving connection. The $\mu=\nu=\rho=0$ point has curvature
\[ R_\nabla s=2 s\wedge t\tens s,\quad R_\nabla t=2 s\wedge t\tens t.\]
We also have an obvious `symmetric lift'  $i(s\wedge t)={1\over 2}( s\tens t+ t\tens s)$ which at the zero parameter point yields 
\[ {\rm Ricci}=s\tens t+t\tens s,\quad S=0\]
but note that $i$ in the $*$-algebra case is not antihermitian in the required sense so that this Ricci is not hermitian.

The above are principal moduli with nontrivial $\sigma$ and zero for the bimodule map $\alpha$ in the general construction; there is also a 4 dimensional moduli of QLCs for $g_2$ with $\sigma=-{\rm flip}$. In general, the space of $*$-preserving `real' pairs $(g,\nabla)$ appears to be generically  7 dimensional, although this remains to be determined. Once known, one could define quantum gravity on $M_2(\C)$ by integration over this moduli, schematically
\[ \<\CO_1\CO_2\>={\int\CD(g,\nabla) e^{\imath {\rm Tr} S[g,\nabla]}\CO_1\CO_2\over \int \CD(g,\nabla)e^{\imath {\rm Tr} S[g,\nabla]} } \]
for the given differential structure and lift $i$ fixed. Here $S[g,\nabla]\in M_2(\C)$ is the Ricci scalar, which generically is expected not to vanish. One could also consider summing or integrating over the choice of differential structure up to diffeomorphism. 

\begin{figure}
\[ \includegraphics[scale=0.5]{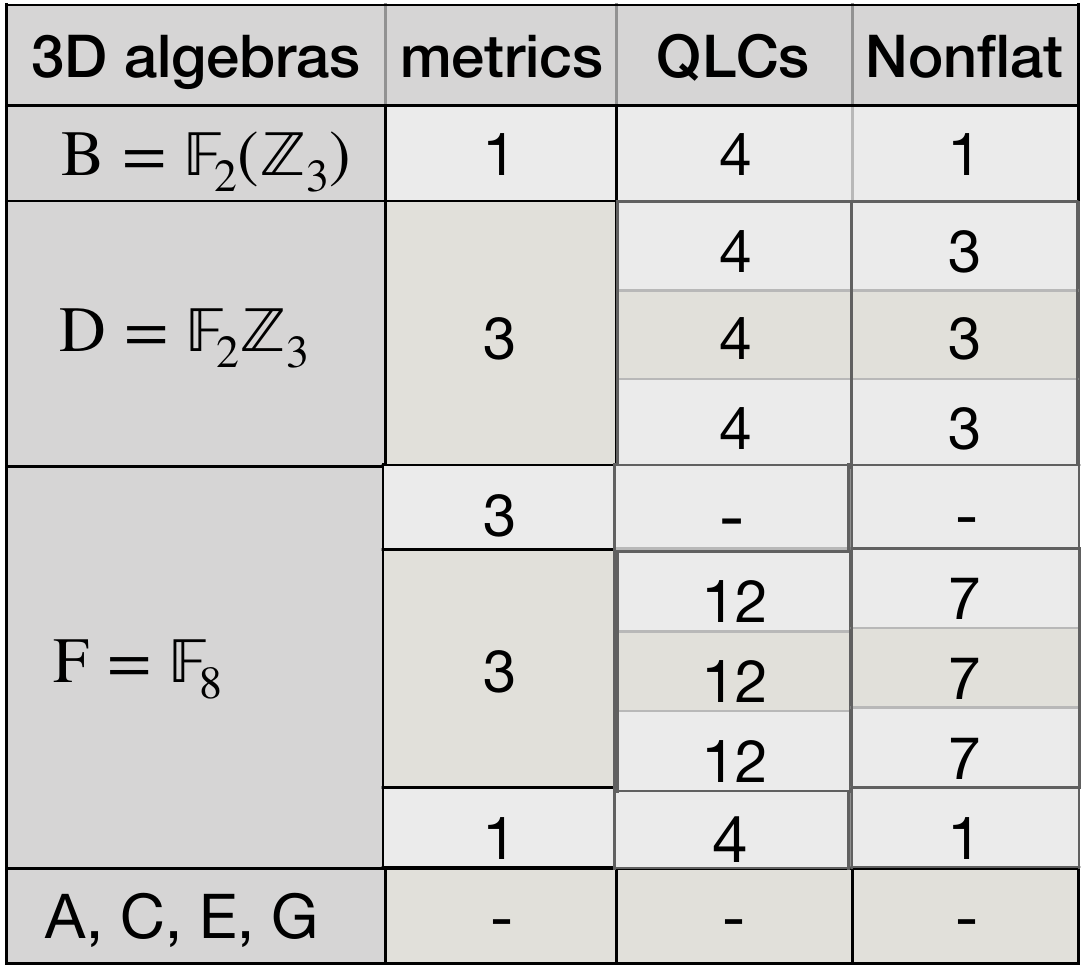}\]
\caption{Number of quantum geometries over $\F_2$ of algebra dimension 3 and 2D parallelisable calculus $\Omega$.  Data from \cite{MaPac2}. 
\label{figtab}}
\end{figure}

Next,  what about the landscape of {\em all} finite quantum geometries of small algebra dimension? This is a tough classification programme and so  far has been achieved\cite{MaPac2} only up to dimension 3 and by specialising to the digital case over the field $\F_2=\{0,1\}$ to simplify the problem. Here algebra dimension 2  has no interesting quantum geometries, while for dimension 3 there are 7 algebras labelled A--G but only B,D,F admit quantum Riemannian geometries as summarised in the table in Figure~\ref{figtab}. Here B is the Boolean algebra of subsets of a set of three elements with calculus and metric corresponding to an equilateral triangle graph, forming the group $\Z_3$. It is a Hopf algebra and its dual is the group algebra D$=\F_2\Z_3$.  Its three quantum metrics are related by an overall element of the algebra  (i.e., `conformally equivalent' in some sense). In each case the 4 QLCs with 1 flat and 3 nonflat is the opposite of what was found for $B$, possibly consistent with  Hopf algebra duality interchanging high and low curvature cf. \cite{Ma:pla,Ma:pri}. For $\F_8$, the 7 metrics are again related by invertible factors but fall into three groups behaving differently as shown in the table. Details of the connections and curvatures are given in \cite{MaPac2}. 

For interesting examples with the algebra noncommutative, we therefore need to go to dimension $n=4$, which was beyond the computer power available when writing \cite{MaPac2} (where QLCs were found by trying some $2^{24}$ possible connection values).  As a result, the landscape of all quantum geometries for $n=4$ is largely unexplored, though some flat connections are known for the algebra $A_2$ in the family of Hopf algebras over $\F_2$ in \cite{BasMa}.  In dimension 4 there are 9 distinct noncommutative algebras\cite{MaPac3} (and 16 commutative ones) over $\F_2$,  and one of the former is of course $M_2(\F_2)$. 

Here we note that  by reducing the above generic solutions for $M_2(\C)$ to the $\F_2$ case we can obtain some, though not necessarily all, of the quantum geometries on $M_2(\F_2)$, at least for the two metrics stated above. Since 2=0 in $\F_2$, we now have $\extd s=\extd t=0$, and indeed $t=\extd E_{12}$, $s=\extd E_{21}$ are exact. Setting the parameters for our generic solutions to all values 0,1 gives us up to 16 and 8 QLCs respectively for our standard metrics, which appear now as
\[ g_1=s\tens t+t\tens s,\quad g_2=s\tens s+ t\tens t.\]
We broadly identify four cases: 

(a) Setting  $\alpha=\beta=\rho=0$ and any $\mu,\nu$ gives the flat zero connection $\nabla s=\nabla t=0$ for both metrics. 

(b) Setting $\alpha=\beta=\rho=1$ and $\mu=\nu=1$ gives  another flat connection
\[ \nabla s=\nabla t= \sigma_1(g_1+g_2),\quad R_\nabla=0\]
for both metrics, on noting that $\nabla (s+t)=0$. Here $\sigma_1=E_{12}+E_{21}$. 

(c) Setting $\alpha=\beta=\rho=1$ and $\mu=\nu=0$ gives another flat connection
\[ \nabla s=E_{12}g_1+ E_{21} g_2,\quad \nabla t=E_{21}g_1+ E_{12} g_2,\quad R_\nabla=0\]
for both metrics,   after some computation. Here $\extd(E_{12}t)=\extd(E_{21}s)=0$ and $\extd(E_{12}s)=\extd(E_{21}t)=s\wedge t$ means that $(\extd\tens\id)\nabla s=(\extd\tens\id)\nabla t=0$. 

\begin{figure}
\[ \includegraphics[scale=0.63]{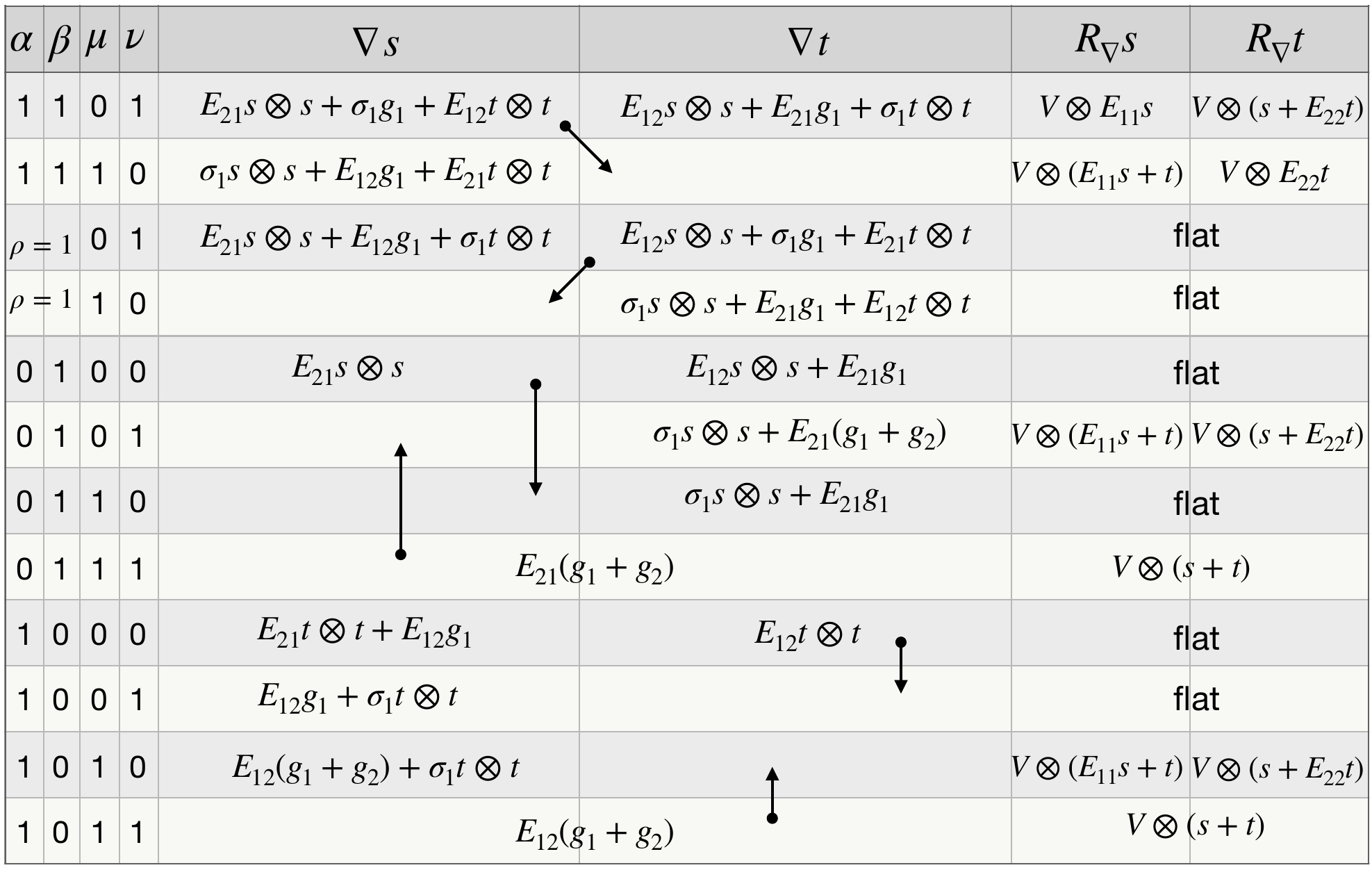}\]
\caption{All QLCs  on $M_2(\F_2)$ arising as limits of connections over $\C$ (which may not be all of them). All are for $g_1$ except the $\rho=1$ rows for $g_2$ which are flat.  We see 6 with curvature, where $V=s\wedge t$ is the `volume form'. 
\label{figtab2}}
\end{figure}

(d) For curvature, we therefore need to look to $\alpha\ne \beta$ or  $\mu\ne \nu$. These are shown in the table Figure~\ref{figtab2} where we see that half of them have Riemann curvature, all QLCs for the metric $g_1$. 

Next, the usual lift $i$ needed for the Ricci curvature, and the Einstein tensor, both present a problem over $\F_2$ in that there is no $1/2$ at our disposal. Our approach  in \cite{MaPac2} is to look at all possible $i$ and tentatively to define ${\rm Eins}={\rm Ricci}+g S$. Now we explore a different idea which is not systematic but applies  when the exterior algebra has suitable form, namely to look for two halves of the classical `antisymmetric' lifts'  separately, without the factor $1/2$. For the $M_2(\F_2)$ calculus as above, obvious choices would be
\[ i_+(s\wedge t)= s\tens t,\quad i_-(s\wedge t)= t\tens s.\]
As a result, there are two natural Ricci tensors ${\rm Ricci}_\pm$ as defined with $i_\pm$. If $S_+=S_-=S$, i.e., if the two Ricci scalars are the same then we propose
\[ {}_2{\rm Ricci}= {\rm Ricci}_++{\rm Ricci}_-,\quad  {}_2{\rm Eins}={}_2{\rm Ricci}+ g S\]
For the 3-dimensional Boolean algebra B, this gives ${}_2{\rm Ricci}=g$, $S=1$ and ${}_2{\rm Eins}=0$ but for the D algebra one has $S_+\ne S_-$ for the obvious lifts (which is not to say the new approach could not work, for suitable $i_\pm$.)

\begin{proposition} On $M_2(\F_2)$ with the metric $g_1$ as above and its two nonflat joint QLCs in Figure~\ref{figtab2} with curvature $R_\nabla s=R_\nabla t=s\wedge t\tens (s+t)$, we have
\[ {\rm Ricci}_+=t\tens (s+t),\quad  {\rm Ricci}_-=s\tens (s+t),\quad S_\pm=1;\quad {}_2{\rm Ricci}=g_1+g_2,\quad {}_2{\rm Eins}=g_2 \]
Moreover, $\nabla$ here are also QLCs for $g_2$ and hence $\nabla( {}_2{\rm Ricci})=\nabla ({}_2{\rm Eins})=0$. 
The three other types of curvature  for QLCs of $g_1$ in Figure~\ref{figtab2} have $S_+\ne S_-$. 
\end{proposition}
\proof We use the curvature as shown and take the `trace' with respect to $g_1$ as this is the relevant metric. Since the output of $R_\nabla$ is the same on $s,t$, 
\[ {\rm Ricci}_+=((s+t ,\ )\tens\id)( s\tens t\tens (s+t))=t\tens (s+t),\quad S_+=1,\]
\[  {\rm Ricci}_-=((s+t ,\ )\tens\id)( t\tens s\tens (s+t))=s\tens (s+t),\quad S_-=1.\]
In this case, we also have ${}_2{\rm Ricci}=(s+t)\tens (s+t)=g_1+g_2$ on adding these. Hence adding $Sg_1$ gives us the other metric. We next look more closely at the two connections in the table with this curvature. The one on the bottom line at $\alpha=\mu=\nu=1$ and $\beta=0$ has braiding
\[\sigma( \begin{cases}s\\ t\end{cases}\nquad\tens s)=s\tens \begin{cases}s\\ t\end{cases}\nquad+ g_1+g_2,\quad \sigma(\begin{cases}s\\ t\end{cases}\nquad\tens t)=t\tens \begin{cases}s\\ t\end{cases}\nquad.\]
Then since $\nabla$ has the same value on $s,t$, 
\begin{align*}\nabla g_2&=\nabla s\tens s+\nabla t\tens t+(\sigma\tens\id)(s\tens \nabla s+ t\tens \nabla t)\\
&=E_{12}(g_1+g_2)\tens (s+t)+ E_{12}(\sigma\tens\id)((s+t)\tens(g_1+g_2))\\
&=E_{12}(s+t)^{\tens 3}+ E_{12}(s+t)^{\tens 3}+ ((\sigma-{\rm flip})\tens \id)((s+t)\tens s\tens (s+ t))=0\end{align*}
where we apply $\sigma$ as ${\rm flip}$ for all cases, plus the additional $\sigma-{\rm flip}$ when acting on $(s+t)\tens s$ in the
second term, which contributes zero. 

The computations for the three other types of curvature of QLCs for $g_1$ are similar. For example, for $R_\nabla s=s\wedge t\tens (E_{11}s+t),\  R_\nabla t=s\wedge t\tens (s+E_{22}t)$, we have
\[ {\rm Ricci}_+=t\tens (t+ E_{11}s),\quad S_+=E_{11},\quad {\rm Ricci}_-=s\tens (s+E_{22}t),\quad S_-=E_{22}\]
so that $S_+\ne S_-$. Interestingly, $S_++S_-=1$.   These same values of $S_\pm$ are also obtained for the other two cases.  Hence for these three curvature types,  our definition of ${}_2{\rm Eins}$ does not apply although for the case shown, we do have ${}_2{\rm Ricci}+S_+t\tens s+S_-s\tens t=g_2$ again. One could still search for other more suitable $i_\pm$  (similarly to the algebra D for $n=3$) and meanwhile, in all cases, we can still use the tentative proposal in \cite{MaPac2} to define ${\rm Eins}_\pm={\rm Ricci}_\pm+ S_\pm g_1$.  \endproof

The Einstein tensor vanishes automatically for a classical 2-manifold,  but this need not be the case in quantum geometry.  We see on this sample of connections on $M_2(\F_2)$ that ${}_2{\rm Eins}$ is conserved when it applies but that the general picture for a suitable Einstein tensor remains inconclusive.  Stepping away from quantum Riemannian geometry towards other applications relevant to computer science,  the  forthcoming work \cite{MaPac3} classifies all quantum groups to dimension 4 over $\F_2$. 

\section{Beyond de Morgan duality}\label{secMor}

Quantum geometry also provides a geometric view of de Morgan duality\cite{Ma:boo}, extending the well-known feature of Boolean algebras that says that the negation of a Boolean expression has the same form with all elements negated, $\emptyset$ and everything swapped, and $\cup,\cap$ swapped. In propositional logic, this sends $a\Rightarrow b$ to the equivalent statement $\bar b\Rightarrow \bar a$, while in terms of the power set $P(X)$ of subsets of a set $X$, the duality sends $a\subseteq X$ to its complement $\bar a$ in $X$. 

This was the topic of my conference talk and I refer to \cite{Ma:boo} for details. In the present notes I instead want to recall the philosophical context and discuss what might come next. Therefore, suffice it to say that one can view $P(X)$ as an algebra over $\F_2$  with product $\cap$ and addition $\oplus$ (the `exclusive or' $a\oplus b=(a\cup b)\cap\overline{a\cap b}$ operation). Next we fix a directed graph on $X$ as vertex set and let ${\rm Arr}$ be the set of arrows. Combining the graph calculus as in Section~\ref{secmarkov} with digital methods as in Section~\ref{secM2}, we set $\Omega^1(P(X))=P({\rm Arr})$ with addition given by `exclusive-or' of subsets of arrows and the noncommutative bimodule structure \cite{Ma:boo} 
\[ a\cap\omega:=\{{\rm arrows\ in\ }\omega\ {\rm with\ tail\ in\ }a\},\quad \omega\cap a:=\{{\rm arrows\ in\ }\omega\ {\rm with\ tip\ in\ }a\}\nonumber\]
\[ \extd a=\{{\rm arrows\ with\ one\ end\ in\ }a\ {\rm and\ other\ end\ in\ }\bar a\}\]
where $a\subseteq X$ and $\bar a$ is its complement. One can go on and define the maximal prolongation exterior algebra $\Omega_{max}(P(X))$ as well as two natural quotients. 

We also define the dual algebra structure $\bar P(X)$ with addition give by `inclusive and' $a\bar\oplus b=(a\cap b) \cup \overline{a\cup b}$  and product $a\cup b$. We define $\Omega^1(\bar P(X))=\bar P({\rm Arr})$ with its `inclusive and'  or `not-exclusive-or' as addition and\cite{Ma:boo} 
\[ a\cup \omega=\{{\rm arrows\ in\ }\omega\ {\rm or\ with\ tail\ in\ }a\},\quad \omega\cup a=\{{\rm arrows\ in\ }\omega\ {\rm or\ with\ tip\ in\ }a\}\nonumber\]
\[ \bar \extd a= \{{\rm arrows\ wholly\ in\  }a\ {\rm or\ wholly\ in\ }\bar a\}=\overline{\extd a}.\]
This too extends to $\Omega_{max}(\bar P(X))$ and its two natural quotients.  

\begin{theorem}\cite{Ma:boo}\label{thdeM}The algebra map  $\bar{\ }: P(X)\to \bar P(X)$ is a diffeomorphism, i.e. extends to a map of the corresponding differential exterior algebras.
\end{theorem}

That $\bar{\ }$ is an algebra isomorphism is the usual de Morgan duality of Boolean algebra in our algebraic language of digital geometry, and the claim is that this extends in the right manner to arrows or differentials of the graph calculus. The extension to 1-forms is just complementation of subsets of arrows. The work \cite{Ma:boo} also shows how these ideas can be extended to any unital algebra $A$ over $\F_2$ with $\bar A$ isomorphic via $\bar a=1+a$ and likewise becoming a diffeomorphism for suitable differential structures. 

This is a purely mathematical result but its philosophical motivation is as follows. Indeed, some 30 years ago in \cite{Ma:pri} I posed the question that if  Boolean algebra are the simplest `theory of physics' then what becomes of de Morgan duality in more advanced theories? I argued that while clearly broken by quantum theory and gravity alone (for example, apples curve space but the presence of not-apples, meaning the absence of applies, does not) such a duality but might re-emerge as a symmetry of quantum gravity. This was and still is meant to be thought-provoking speculation rather than something understood, but the idea was that {\em we} might say that a region of space is `as full of apples' as GR allows (forming a black hole and expanding if we put more apples in) while someone else using the dual picture might say that this same region of space was as empty of not-apples as their quantum field theory allows (where in QFT space is never completely empty in some sense due to vacuum fluctuations). I don't know if this vision is achievable but what we can do is move quantum gravity down to the digital level of Boolean algebras and see if de Morgan duality indeed holds there. This is what have now found in \cite{Ma:boo} at the Boolean level in so far as GR is a theory of metrics and connections on $\Omega^1$;  all of that works as we have seen in a dual version. We also needed an element of `quantum' in that our extension of $\cap,\cup$ to arrows was noncommutative, then Theorem~\ref{thdeM} says 
that de Morgan duality indeed holds as part of general covariance in some extended sense. 

This sense is admittedly a little wierd. In usual GR a diffeomorphism is induced by a underlying set map, but complementation is different and operates directly on subsets of $X$. For example, if $S\in P(X)$ is the Ricci scalar for a connection then for the same geometry it appears as $\bar S$ on the other side of de Morgan duality. So in a region where the curvature characteristic function has value 1 for us, it has value 0 for them. Similarly, the inner element $\theta\in \Omega^1(P(X))$ is the sum of all arrows, or in some sense the `maximal density' differential form. Its role on the de Morgan dual side is played by the zero element which is inner for $\Omega^1(\bar P(X))$ but from the first point of view is literally the zero differential form. All of this is suggestive of the black-hole/vacuum discussion above but is probably the most we can say at the Boolean level (where it is hard to think about actual quantum theory). 

On the other hand, more advanced theories of physics could still have the duality we seek in an increasingly visible form. Thus, speculatively, just as Schroedinger's cat is in a mixed state that is {\em neither dead or alive}, I have proposed \cite{Ma:qg3}  {\em co-Schroedinger's cat} as a cat falling into a black hole. This is {\em both dead} in finite proper time {\em and alive} forever in the frame of the observer at infinity. In other words, while quantum theory is intuitionistic as in a Heyting algebra, where we relax the rule that $a\cup\bar a=$everything,  gravity might be expected to be cointuitionistic in character in the de Morgan dual sense, as in a coHeyting algebra where we relax the rule that $a\cap\bar a=\emptyset$. The latter has also been proposed for other reasons in \cite{Law}  as geometric in nature with $\del a=a\cap \bar a$ a kind of boundary of $a$. This then requires both effects or quantum-and-gravity for the symmetry to be maintained. This suggests: 

\begin{quote} Can we extend quantum differential geometry to Heyting and coHeyting algebras and thereby extend the negation duality to a diffeomorphism? 
\end{quote}

The problem is that a Heyting algebra has both a `meet' product $\cap$ and a `join'  $\cup$ but the latter does not in general provide an  addition rule and does not play well with negation in order to be able to define a more suitable $\oplus$ over which $\cap$ distributes, i.e. we do not in general have the basics for an actual algebra in the sense of a ring (before even considering a field).  I do not know the answer but I would like to suggest an approach. In fact both Boolean algebras $P(X)$ and Heyting algebras (and a lot more) are examples of preorders $(P,\le)$ by which we mean an extended directed graph (adding in all self-arrows) which is closed under composition of arrows (here $a\le b$ corresponds to an arrow $a\to b$ and in the Boolean case is just subset inclusion). Both cases actually have rather more structure, namely min and max elements $0,1$ (the empty set and the whole set in the Boolean case), unital tensor product $\tens$ (the $\cap$) and an `internal hom' $\inthom$ leading to a `dual' $\bar a=\inthom(a,0)$ (the negation) making the preorder closed symmetric monoidal in the sense of \cite{ApCat} (this is a baby closed symmetric monoidal category in the sense that there is at most one morphism between any two objects). Abstractly, $\inthom$ is a binary operation characterised by
 $a\tens b\le c$ if and only of $a\le \inthom(b,c)$, which you can check is true in the Boolean case with $\inthom(b,c)=\bar b\cup c$  (and its existence is a definition in the Heyting case). In this sense de Morgan duality has a similar conceptual flavour to vector space duality in the tensor category of vector spaces, which is in the right direction towards connecting it with observable-state or Hopf algebra duality\cite{Ma:sel}. 

Next,  instead of working directly with the Boolean or Heyting algebra $P$ as our `algebra of functions', we work now with functionals, i.e. the algebra $A=\C(P)$ of functions on $P$.  As $P$ being a preorder has a directed graph structure, we have
\[ \extd \Phi= \sum_{a\le b} (\Phi(b)-\Phi(a))\omega_{a\le b}\]
for all functionals $\Phi\in \C(P)$ as a graph calculus on $P$. The Boolean or Heyting negation as a set map $P\to P$ induces an algebra map on $\C(P)$ and this again is differentiable. We do not directly see the `manifold' structure on $X$ in the Boolean case where $P=P(X)$ and $X$ itself has a directed graph with arrow set {\rm Arr}. However, we can similarly look at functionals of the differentials on $X$ understood now as functions on $P({\rm Arr})$. We therefore have the ingredients for some form of variational calculus for functionals of functions on $X$ and their derivatives,  in which  the graph differentials on $X$ enter, much as quantum field theory on the space of functions on a manifold still captures information about the manifold. When $P$ is now some Heyting algebra or indeed any preorder, we can look for similar structures at the level of $\C(P)$ with a calculus on $P$ handled implicitly. Given the importance of Heyting algebras and topos theory in physics, see e.g. \cite{DorIsh}, it should be useful to explore their quantum Riemannian geometry, extending what should be achievable in the Boolean case, even if we have to handle it implicitly in terms of functionals. What may emerge is something like working over an algebra but in terms of $\cup$ for addition, or some kind of $\oplus$ but with weakened distributivity. Some ideas as to the latter were previously in \cite[Sec.~7]{Ma:sel}.  

Heyting algebras provide the link up with probability as follows. We consider, as we did in Section~\ref{secmarkov}, functions on $X$ now with values in $[0,1]$, i.e. probabilities. $[0, 1]$ itself is a Heyting algebra and $C(X,[0,1])$ inherits its features pointwise, including a preorder (where $f\le g$ if $f(x)\le g(x)$ for all $x\in X$), distributive, commutative and associative meet and join operations $(f\cup g)(x)={\rm max}(f(x),g(x))$ and $(f\cap g)(x)={\rm min}(f(x),g(x))$ as well as an internal hom $\inthom(f,g)(x)=1$ wherever $f(x)\le g(x)$, and $g(x)$ at other points. The negation $\hom(f,0)$ here is $1$ at the zeros of $f$ and otherwise $0$, which still interchanges $\cup$ and $\cap$ as for de Morgan duality but which loses information and hence is not a proper duality (indeed, double negation is not the identity except on the Boolean subalgebra with values in $\{0,1\}$). By contrast, we observe that there is a different complementation which does square to the identity, 
\begin{equation}\label{barprob} \Bar f(x)=1-f(x),\quad\forall x\in X,\end{equation}
and which again still interchanges $\cap,\cup$ as expected for de Morgan duality. It sends high probability to low probability in the same spirit as discussed in the Boolean case (as well as in the spirit of the prophetic work of the English satirist Douglas Adams). Moreover,  we can  still take the preorder `field theory' point of view and do differential geometry and perhaps variational calculus on $A=\C(C(X,[0,1]))$ as above, but for the geometry of $C(X,[0,1])$ itself we again do not have an algebra due to lack of a proper addition.  

In fact, $[0,1]$ and $C(X,[0,1])$ have another product which is the more obvious one given by the usual product in $[0,1]$ as real numbers and the ordinary pointwise product $fg$ of functions with values in $[0,1]$. We can then generate its `de Morgan dual' addition via the complementation (\ref{barprob}) as a new operation which we denote $f\oplus g=1-(1-f)(1-g)=f+g -fg$ in the spirit of \cite{Ma:boo} for the $\F_2$ case. This is associative and has a zero but one has $f(g\oplus h)\le fg\oplus fh$ rather than full distributivity, so this pair of operations again does not quite give an algebra. 
This product in $\R$ again makes both $[0,1]$ and $C(X,[0,1])$ closed symmetric monoidal preorders, but different from our previous ones using $\cap$. The internal hom in $[0,1]$ is $\inthom(p,q)=1$ if $p\le q$ else $q/p$ and similarly for functions $\inthom(f,g)(x)=1$ at points where $f(x)\le g(x)$ and $g(x)/f(x)$ at other points. In fact the negation $\inthom(\ ,0)$ here is the same as before (but the internal hom's in general are different).  

All four binary operations on $C(X,[0,1])$ remain tightly related and we can consider the usual pointwise $fg$ product along with $f\cup g$ given by the pointwise maximum. This has distributivity $f(g\cup h)=(fg)\cup (fh)$ but, as mentioned, $\cup$ is not an addition law. On the other hand, if we now replace probabilities in $[0,1]$ by minus their logarithm as we did in Section~\ref{secmarkov} in discussing geodesic paths, then we equivalently have functions on $X$ with values in $\R_{\ge 0}$ with pointwise addition and minimum, i.e. in the tropical version of $\R$. This point of view has, for example,  applications in statistical inference\cite{Sturm}.  So one may possibly garner ideas for differential geometry on $X$ itself from tropical algebraic geometry. From probability functions to positive linear functionals on noncommutative algebras is a further but well-known step. In this way, we have sketched a path from Boolean algebras or logic to probability to full quantum geometry over $\C$, with some kind of generalised de Morgan duality playing a pivotal role. 

Finally, it should be mentioned that at the time of \cite{Ma:pri}, such duality ideas motivated the view that quantum gravity needs geometry that is at the same time quantum or noncommutative, with the duality realised slightly differently in concrete `toy models'\cite{Ma:pla} as observer-observed, representation theoretic and Hopf algebra duality. The bicrossproduct quantum groups associated to Lie group factorisations emerging from this, as well as the Drinfeld-Jimbo one q-deforming complex simple Lie groups, contributed to a concrete `constructive' approach to such quantum Riemannian geometry and included the first convincing model \cite{MaRue} of quantum spacetime with quantum symmetry. These ideas are also tied up with quantum group Fourier transform and in physical terms with `Born reciprocity' and were at the root of my proposal back in \cite{Ma:pla,Ma:pri} as well in later works\cite{Ma:ess, Ma:qg3, Ma:sel,Ma:eme}.  Although now somewhat established, the  deepest aspect of this duality -- swapping observables and states -- remains unexplored and should relate to issues of measurement, probability and logic much as above, for example to the entropic arrow of time\cite{Ma:ran}. How exactly it relates to a duality growing out of de Morgan duality remains a topic for further thought, albeit the notion of bi-Heyting algebra \cite{ReyZol} could be a step in this direction.

\section{Concluding remarks} \label{secrem}

The main new results of the paper are our quantum geometric view of Markov processes and the new concept of an underlying `Schroedinger process' in Section~\ref{secmarkov}. We conclude with some comments on these. 

Our first comment is that it could potentially be interesting to extend the graph calculus used here to quivers (where there can be self-arrows {\em and} multiple arrows between vertices). We have already seen the need for self-arrows but now we can go one step further to multiple arrows. One still has a differential calculus in the sense of part of a DGA but  not all 1-forms need be sums of elements of the form $a\extd b$\cite{MaTao:dua}. One can still do quantum Riemannian geometry but now a metric is not a number on each edge but a matrix \cite{MaTao:dua}. In the Markov case we could imagine a completely positive matrix, though this remains to be established. The goal would be to generalise Markov processes to allow different flavours of transition eg due to different types of processes between vertices. We also note \cite{Sturm} concerning the use of graphs with arrows labelled by linear forms to describe statistical models. Moreover, transitively closed extended graphs (i.e., preorders) are baby versions of categories and one could ask if differential geometry could generalise further to this categories in the spirit of Section~\ref{secMor}.

Our second comment concerns the notion of geodesics. In noncommutative geometry one does not have points, so nor does one have geodesics as paths. Instead we have to work directly with functions and our first thought might be probability density functions. However, quantum Riemannian geometry in the constructive form \cite{BegMa} is formulated in a linear setting so one is led to `amplitudes' $\psi$ in a Hilbert space evolving in time and probability density $f=|\psi|^2$. This was the motivation behind \cite{Beg:geo,BegMa:geo} and indeed quantum mechanics seems to be tightly linked with this point of view. This suggests that the philosophy of quantum mechanics, the measurement problem and so forth might be clearer as geodesic flow of some kind in quantum geometry. This, and more generally the role of quantum geometry (which is about extending macroscopic concepts to the quantum level) in the nature of measurement is a topic for further study. 

Indeed, the concrete result in the present paper is that if one goes further and looks at discrete-time Schroedinger processes in discrete quantum geometry, then coming out of the discreteness is a correction to the familiar equation $\dot f=-\nabla\cdot J$  in usual quantum mechanics in which there is an extra Markov process induced on $f$. This has more of a classical flavour and could play a role in measurement collapse. Such finite processes should also be of interest in quantum computing, where finite-dimensional unitary  matrices are `gates' and we have shown how they could be constructed quantum geometrically.

\end{document}